\documentclass[twocolumn,prl,showpacs,amsmath,amssymb]{revtex4}

\usepackage{graphicx}
\usepackage{dcolumn}
\usepackage{bm}

\begin{document}

\title{Experimental determination of the evolution of the Bjorken 
integral at low $Q^2$}

\author{
A.~Deur$^{\njlab}$,
P.~Bosted$^{\njlab , \numass}$,
V.~Burkert$^{\njlab}$,
G.~Cates$^{\nuva}$,
J.-P.~Chen$^{\njlab}$,
Seonho~Choi$^{\ntemple}$, 
D.~Crabb$^{\nuva}$,
C.W.~de~Jager$^{\njlab}$, 
R.~De~Vita$^{\ninfn}$, 
G.~E.~Dodge$^{\nodu}$, 
R.~Fatemi$^{\nuva}$, 
T.~A.~Forest$^{\nltec}$, 
F.~Garibaldi$^{\ninfns}$,
R.~Gilman$^{\njlab , \nrutgers}$,
E.~W.~Hughes$^{\ncaltech}$, 
X.~Jiang$^{\nrutgers}$,
W.~Korsch$^{\nuk}$, 
S.~E.~Kuhn$^{\nodu}$, 
W.~Melnitchouk$^{\njlab}$,
Z.-E.~Meziani$^{\ntemple}$, 
R.~Minehart$^{\nuva}$, 
A.~V.~Skabelin$^{\nmit}$,  
K.~Slifer$^{\nuva , \ntemple}$, 
M.~Taiuti$^{\ninfn}$,
J.~Yun$^{\nodu}$}

\affiliation{
\baselineskip 2 pt
\centerline{{$^{\njlab}$Thomas Jefferson National Accelerator Facility, Newport News, VA 23606}}
\centerline{{$^{\numass}$University of Massachusetts, Amherst MA 01002}}
\centerline{{$^{\nuva}$University of Virginia, Charlottesville, VA 22904}}
\centerline{{$^{\ntemple}$Temple University, Philadelphia, PA 19122}}
\centerline{{$^{\ninfn}$INFN, Sezione di Genova , 16146 Genova, Italy}}
\centerline{{$^{\nodu}$Old Dominion University,  Norfolk, VA 23529}}
\centerline{{$^{\nltec}$Louisiana Tech University, Ruston, LA 71272}}
\centerline{{$^{\ninfns}$ISS/INFN Roma1, gr. coll. Sanita', Viale Regina Elena 299, 00161 Rome, Italy}}
\centerline{{$^{\nrutgers}$Rutgers University, Piscataway, NJ 08854}}
\centerline{{$^{\ncaltech}$California Institute of Technology, Pasadena, CA 91125}}
\centerline{{$^{\nuk}$University of Kentucky, Lexington, KY 40506}}
\centerline{{$^{\nmit}$Massachusetts Institute of Technology, Cambridge, MA 02139}}
}

\newcommand{\ninfn}{5}
\newcommand{\nuk}{11}
\newcommand{\nmit}{12}
\newcommand{\numass}{2}
\newcommand{\nltec}{7}
\newcommand{\nodu}{6}
\newcommand{\ntemple}{4}
\newcommand{\njlab}{1}
\newcommand{\nuva}{3}
\newcommand{\nrutgers}{9}
\newcommand{\ncaltech}{10}
\newcommand{\ninfns}{8}

\date{\today}

\begin{abstract}
We extract the Bjorken integral $\Gamma^{p-n}_1$ in the range 
$0.17 < Q^2 < 1.10$~GeV$^2$
from inclusive scattering of polarized electrons by polarized
protons, deuterons and $^3$He, for the region in which the integral is
dominated by nucleon resonances.
These data bridge the domains of the hadronic and partonic descriptions
of the nucleon.
In combination with earlier measurements at higher $Q^2$, we extract
the non-singlet twist-4 matrix element $f_2$.
\end{abstract}


\maketitle

For almost 50 years experimental and theoretical research efforts in
hadronic physics have sought to understand the structure of the nucleon.
With the development of Quantum Chromodynamics (QCD), these studies have
focused on obtaining an accurate description of nucleon structure
in terms of fundamental quark and gluon degrees of freedom.
A powerful tool has been deep inelastic lepton scattering
from nucleons and nuclei, and the associated theoretical machinery of 
the operator
product expansion (OPE), which allows the interpretation of the measured 
structure functions in terms of parton momentum and spin distribution 
functions.

Experiments using polarized beams and targets have played a
critical role in testing the application of QCD to nucleon
structure~\cite{JI-filippone}.
The Bjorken sum rule~\cite{Bjorken}, which relates the first moment
of polarized deep inelastic structure functions to nucleon ground state 
properties, has been an important part of these studies.
At infinite four-momentum transfer squared, $Q^2$, the sum rule 
reads
\begin{equation}
\Gamma_1^{p-n} \equiv \Gamma_1^p - \Gamma_1^n
\equiv \int_0^1 dx \left( g_1^p(x) - g_1^n(x) \right)
= \frac{g_A}{6},
\label{eq:bj}
\end{equation}
where $g_1^p$ and $g_1^n$ are the spin-dependent proton and neutron
structure functions, respectively. Here, $g_A$ is the nucleon axial 
charge and $x = Q^2/2M\nu$, with $\nu$ the energy transfer and
$M$ the nucleon mass. 
The sum rule has been verified experimentally to better than 10\%
\cite{SLAC,E155,SMC}.

Because the Bjorken sum rule relates differences of proton and
neutron structure function moments, $\Gamma_1^p$ and $\Gamma_1^n$,
only flavor non-singlet quark operators appear in the OPE.
Another simplification arises at low $Q^2$ where the resonance contributions
to the proton and neutron, in particular that 
of the $\Delta(1232)$ resonance, partly cancel.
This cancellation simplifies calculations based on chiral
perturbation theory ($\chi$PT), and may extend the $Q^2$ range of
their applicability. The gap between the domains of validity
for perturbative QCD (pQCD) and $\chi$PT might even be bridged, 
enabling for the first time a fundamental theoretical description of nucleon
structure from large to small scales~\cite{Volker}.
The Bjorken sum rule is therefore 
relevant for understanding the transition from pQCD to nonperturbative
QCD.

In this Letter we report on a determination of the Bjorken integral using
data obtained at the Thomas Jefferson National Accelerator Facility
(Jefferson Lab) over the $Q^2$ range of 0.17--1.10~GeV$^2$.
Combined with higher $Q^2$ data from earlier experiments, we analyze
the data using the OPE and extract the $1/Q^2$ higher twist
corrections to the integral at intermediate values of $Q^2$.

The data were obtained in three different experiments
using polarized electrons on polarized proton~\cite{eg1a proton},
deuterium~\cite{eg1a deuteron} and $^3$He~\cite{E94010-1,E94010-2}
targets.
To analyze the scattered electrons, the proton and deuteron 
experiments used the CEBAF Large Acceptance Spectrometer (CLAS) 
in Hall B~\cite{Hall B nim}, while the $^3$He experiment used 
the two High Resolution Spectrometers in Hall A~\cite{HallA nim}.

The individual measurements of the proton, neutron and deuteron
integrals $\Gamma_1^{p,n,d}$ have been
reported elsewhere~\cite{eg1a proton,eg1a deuteron,E94010-1,E94010-2}.
To form the isovector combination $\Gamma_1^{p-n}$ we subtract from the
experimental values of $g_1^p$ the values of $g_1^n$ extracted from the
$^3$He or the deuteron measurements.
However, in order to combine these data, the $Q^2$ values
at which $g_1^p$ and $g_1^n$ were obtained must coincide.
We chose to re-analyze the $^3$He data at six values of $Q^2$ which
match the ones of the proton data and differ from the values reported in
Refs.~\cite{E94010-1,E94010-2}.
For the deuteron measurement, given the larger uncertainties, we simply
interpolated the proton data points to match the four $Q^2$ points
of the deuteron data.
The additional systematic uncertainty from the interpolation is
negligible.

\begin{figure}[ht!]
\begin{center}
\centerline{\includegraphics[scale=0.38, angle=0]{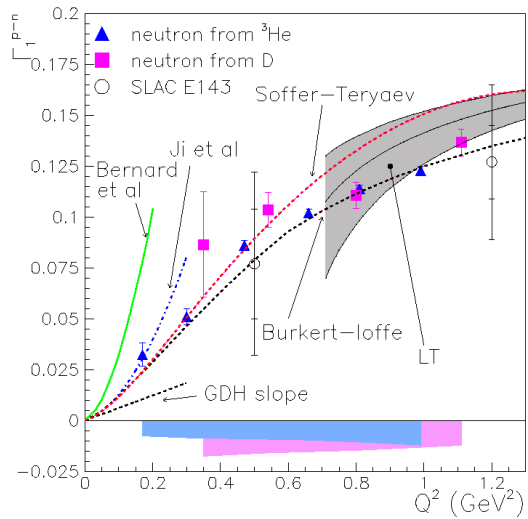}}
\end{center}
\caption{Inelastic contribution to the Bjorken sum.
	The triangles (squares) represent the results with the neutron 
        extracted from $^3$He (deuteron) data, with the horizontal bands
	covering the 0.17-0.99~GeV$^2$ and 0.34-1.1~GeV$^2$ ranges
	the corresponding systematic uncertainties.
	The E143 data are shown for comparison.
	The gray band represents the leading-twist (LT) contribution
	calculated to third order in $\alpha_s$.
	The curves correspond to theoretical calculations (see text).}
\label{fig:bjs}
\end{figure}

The three experiments~\cite{eg1a proton,eg1a deuteron,E94010-1,E94010-2}
have measured $g_1$ up to an invariant mass $W=2$~GeV. The unmeasured 
contributions to the proton and neutron integrals, corresponding to the
low-$x$ domain,  need to be consistently accounted for.
In the current analysis, the fit from Bianchi and Thomas 
\cite{Bianchi} was used to estimate the low-$x$ contribution to the 
moments up to $W^{2}=1000$~GeV$^2$. The uncertainty on this
contribution was evaluated by taking the quadratic sum of the 
differences induced by independently varying each parameter of the fit 
within the range of values given in \cite{Bianchi}. The contribution beyond 
1000 GeV$^2$ was determined using a Regge parametrization constrained 
so that the Bjorken integral at $Q^{2}=5$~GeV$^2$, from the world data 
complemented by the Bianchi and Thomas fit and our Regge parametrization, 
agrees with the sum rule.
The systematic uncertainties from the neutron and proton data have
been added in quadrature.
The moment $\Gamma_1^n$ was extracted from $^3$He or deuterium data
using the formalism of nucleon effective polarizations
\cite{degli atti,extract n from d}.
The resulting $\Gamma_1^{p-n}$ is shown in Fig.~\ref{fig:bjs}
by the filled symbols, with the values given in Table~I. Note 
that only the inelastic contributions are included in $\Gamma_1^{p-n}$.
Data from the SLAC E143 experiment~\cite{e143} are also shown 
(open circles) for 
comparison.

\begin{table}
\caption{Inelastic contributions to the Bjorken sum. 
The second and third columns
give the sum and its uncertainty for $W<2$ GeV. The fourth to sixth columns 
give the total sum and its uncertainties. The last column indicates
the origin of the neutron information (from $^3$He or from deuteron data).}
\small{
\begin{tabular}{|c|c|c|c|c|c|c|}
\hline 
$Q^2$ (GeV$^2$) &
$\Gamma_{1(res)}^{p-n}$&
$\sigma_{(res)}^{syst}$&
$\Gamma_{1(Tot)}^{p-n}$&
$\sigma_{(Tot)}^{syst}$&
$\sigma^{stat}$&
$n$ \tabularnewline
\hline
\hline 
0.17&
0.0134&
0.0073&
0.0326&
0.0076&
0.0057&
$^{3}$He\tabularnewline
\hline 
0.30&
0.0181&
0.0079&
0.0510&
0.0085&
0.0039&
$^{3}$He\tabularnewline
\hline 
0.34&
0.0498&
0.0165&
0.0864&
0.0202&
0.0266&
$\rm D$\tabularnewline
\hline 
0.47&
0.0381&
0.0071&
0.0860&
0.0089&
0.0025&
$^{3}$He\tabularnewline
\hline 
0.53&
0.0507&
0.0121&
0.1035&
0.0170&
0.0095&
$\rm D$\tabularnewline
\hline 
0.66&
0.0394&
0.0058&
0.1019&
0.0095&
0.0020&
$^{3}$He\tabularnewline
\hline 
0.79&
0.0395&
0.0122&
0.1107&
0.0176&
0.0076&
$\rm D$\tabularnewline
\hline 
0.81&
0.0413&
0.0056&
0.1138&
0.0109&
0.0019&
$^{3}$He\tabularnewline
\hline 
0.99&
0.0400&
0.0049&
0.1229&
0.0120&
0.0019&
$^{3}$He\tabularnewline
\hline 
1.10&
0.0477&
0.0084&
0.1366&
0.0166&
0.0076&
$\rm D$\tabularnewline
\hline
\end{tabular}
}
\end{table}

The data are compared with theoretical calculations based on $\chi$PT
and with phenomenological models. At $Q^{2}=0$, the slope of the Bjorken 
integral is constrained by the Gerasimov-Drell-Hearn (GDH) sum rule
~\cite{GDH,Ji-GDH}.
The Soffer--Teryaev model~\cite{Soffer} agrees only with the low-$Q^2$
data. The overestimate at larger $Q^2$ was traced back to 
the QCD radiative corrections and has now been corrected~\cite{Soffer2}. 
The Burkert--Ioffe model~\cite{Burkert and Ioffe} agrees well 
with the data over the full range covered in Fig.~1. This may indicate 
that vector meson dominance gives a reasonable picture of the 
parton-hadron transition. At low $Q^2$ attempts have also been made to 
calculate $\Gamma_1^{p-n}$ using $\chi$PT~\cite{meissner chipt,ji chipt}.
The calculation done in the heavy baryon approximation~\cite{ji chipt} 
seems to agree better with the data. 
At higher $Q^2$ the data are compared with the leading-twist
calculation (gray band in Fig.~1), which corresponds to incoherent
scattering from individual quarks.
In pQCD, gluon radiation causes scaling violations in structure
functions, and introduces an $\alpha_s$ dependence on the right hand side
of Eq.~(\ref{eq:bj}).
At leading-twist, the pQCD result at third order in $\alpha_s$
(in the $\overline{\textrm{MS}}$ scheme) is
\begin{eqnarray}
\Gamma_1^{p-n}=\frac{g_A}{6}
  \left[ 1-\frac{\alpha_s}{\pi}
	  -3.58\left(\frac{\alpha_s}{\pi}\right)^2
	  -20.21\left(\frac{\alpha_s}{\pi}\right)^3
  \right].	\nonumber\\
& &
\label{eq:gamLT}
\end{eqnarray}
The gray band in Fig.~1 represents the uncertainty in $\Gamma_1^{p-n}$
due to the uncertainty in $\alpha_s$. There is reasonable agreement
between the leading-twist prediction and the data. Their difference is
related to higher twist effects that should become important at low $Q^{2}$.
In particular, application of the OPE to moments of structure
functions requires the expansion of the {\em total} moment rather 
than the {\em inelastic} moment as in Fig.~1.
While the elastic contribution is negligible at high $Q^2$,
it dominates at low $Q^2$.
Fig.~2 shows the total moment, including the elastic contribution,
calculated from the form factor parameterizations from
Ref.~\cite{Mergell}. 
In addition to the JLab data, we also plot data at higher $Q^2$
from the SLAC E143~\cite{E143-2} and E155~\cite{E155}, DESY 
HERMES~\cite{HERMES} and CERN SMC \cite{SMC} experiments.
For consistency, the low-$x$ contributions, outside of the measured
regions, have been re-evaluated using the same procedure as described earlier.

The OPE analysis allows one to expand the total moment $\Gamma_1^{p-n}$ 
in powers of $1/Q^2$:
\begin{eqnarray}
\Gamma_1^{p-n}
&=& \sum_{i=1}^\infty {\mu^{p-n}_{2i} \over Q^{2i-2}},
\end{eqnarray}
where the leading twist $i=1$ coefficient is given in
Eq.~(\ref{eq:gamLT}).
The coefficients $\mu^{p-n}_{2i}$ for $i>1$ represent matrix elements 
of higher twist operators. The matrix elements contain information on the long
range, nonperturbative interactions or correlations between partons. 
In particular, the $1/Q^2$ correction term is 
\cite{shuryak,ji2}
\begin{eqnarray}
\mu_4^{p-n}
&=& \frac{M^2}{9}
    \left( a^{p-n}_2 + 4 d^{p-n}_2 + 4 f^{p-n}_2 \right),
\end {eqnarray}
where $a_2^{p-n}$ is the target mass correction given by the
$x^2$-weighted moment of the leading-twist $g_1$ structure function,
and $d_2^{p-n}$ is a twist-3 matrix element given by
\begin{eqnarray}
d^{p-n}_2
&=& \int_0^1 dx~x^2 \left( 2g^{p-n}_1 + 3g^{p-n}_2 \right).
\end{eqnarray}
The twist-4 contribution, $f_2^{p-n}$, given by a mixed quark-gluon
operator, is related to the color electric and magnetic
polarizabilities of the nucleon~\cite{MANK,JI_CHI}.

\begin{figure}[ht!]
\begin{center}
\centerline{\includegraphics[scale=0.38, angle=0]{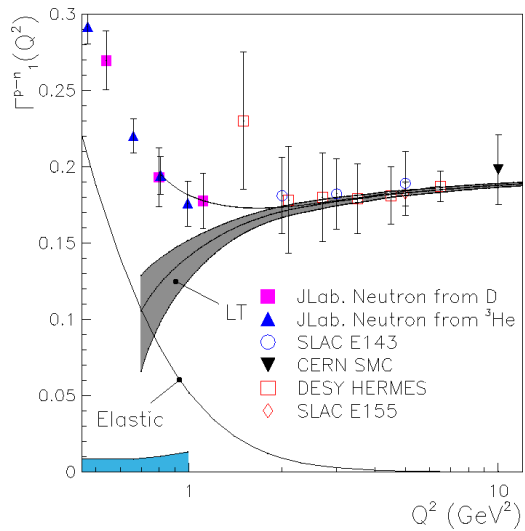}}
\end{center}
\caption{Total moment $\Gamma_1^{p-n}$, including the inelastic
	contribution from Fig.~1 together with the elastic.
	The data extracted from $^3$He (deuterium) together with
	proton data are indicated by the triangles (squares).
	The leading-twist (LT) contribution is given by the
	gray band. The point to point correlated uncertainty
	for the data extracted from $^3$He and proton is shown
	by the horizontal band. The error bars on the symbols represent
	the uncorrelated uncertainty. The data from other experiments
	are assumed to be uncorrelated.
	The fits to the total moment are indicated by the solid
	curves.}
\label{fig:ht}
\end{figure}

The $a_2^{p-n}$ correction, which is twist-2, is
calculated using the fit of polarized parton distributions from
Ref.~\cite{BB}.
The $d_2^{p-n}$ matrix element is obtained from~\cite{a1n}.
With these inputs, the data on $\Gamma_1^{p-n}$ in Fig.~2 can be
used to extract $f_2^{p-n}$. As an additional parameter in the fit, 
we include the $1/Q^4$ coefficient
$\mu_6^{p-n}$.  For the leading twist contribution, we constrain the
low-$x$ extrapolation by assuming the validity of the Bjorken sum rule
for $Q^2 > 5$~GeV$^2$.  In fact, our low-$x$ extrapolation gives
$g_A^{\rm fit} = 1.270 \pm 0.045$, which is very close to the empirical
value $g_A = 1.267 \pm 0.004$.  The higher twist terms are then determined 
from the
$Q^2 < 5$~GeV$^2$ data using our fitted value of $g_A$.
The point to point correlated uncertainty for the JLab data extracted
from $^3$He and hydrogen has been disentangled from the uncorrelated
uncertainty, and only the latter is used in the fit.  The correlated
systematics are then propagated to the fit result, as is the uncertainty
arising from $\alpha_s$.  The data from the other experiments are treated
as uncorrelated from point to point.
                                                                               
It is not clear {\em a priori} over which $Q^2$ range
the $1/Q^2$ expansion should be valid.
For instance, at $Q^2 \approx 0.7$~GeV$^2$ the elastic and leading twist
contributions are of comparable magnitude.
  Fitting over the range $0.8 < Q^2 < 10$~GeV$^2$ gives
$f_2^{p-n} = -0.13 \pm 0.15 ({\rm uncor})^{+ 0.04}_{-0.03}({\rm cor})$,
normalized at $Q^2=1$~GeV$^2$, where the first and second errors are
uncorrelated and correlated, respectively, and $\mu_6^{p-n} /M^4 
= 0.09 \pm 0.06({\rm uncor}) \pm 0.01({\rm cor})$. The contribution to
the total uncertainty from the elastic form factors is negligible.
Starting the fit at a lower $Q^2$, $Q^2 = 0.66$~GeV$^2$, yields the
more negative value
$f_2^{p-n} = -0.18 \pm 0.05 ({\rm uncor})^{+0.04}_{-0.05}({\rm cor})$,
and a larger value for $\mu_6^{p-n}$,
$\mu_6^{p-n}/M^4 = 0.12 \pm 0.02 ({\rm uncor}) \pm 0.01 ({\rm cor})$,
with somewhat smaller errors.  The results of the two fits are shown in
Fig. 2, but are almost indistinguishable. 
At $Q^{2}=1$~GeV$^2$, the
$1/Q^4$ contribution is $\mu_6^{p-n}/Q^4 \simeq 0.09 \pm 0.02$,
which is of similar magnitude and of opposite sign to the $1/Q^2$ term,
$\mu_4^{p-n}/Q^{2} \simeq -0.06 \pm 0.02$, obtained by adding the extracted
$f_2^{p-n}$ value to $d_2^{p-n}$ and $a_2^{p-n}$ in Eq.~(4).  This may
explain why the leading twist description agrees well with the data down
to surprisingly small values of $Q^2$ ($\sim 1$~GeV$^2$), and could be a
hint that quark-hadron duality might work well for $p-n$ non-singlet
quantities.

These results also suggest that at these lower $Q^2$ values the twist
expansion may not converge very quickly, and that higher twist terms may
be needed.
Including a $\mu_8^{p-n}/Q^6$ term, however, gives significantly larger
uncertainties on the higher twist contributions, making them compatible
with zero.  Starting the fit at $Q^2 = 0.47$~GeV$^2$, for instance, gives
$f_2^{p-n} = -0.14 \pm 0.10(\rm uncor) \pm +0.04(\rm cor)$,
$\mu^{p-n}_6/M^4 = 0.09 \pm 0.08(\rm uncor)^{+0.03}_{-0.04}(\rm cor)$ and 
$\mu^{p-n}_8/M^6 = 0.01 \pm 0.03(\rm uncor) \pm 0.02(\rm cor)$.

The results for $f_2^{p-n}$ can be  
compared to non-perturbative model predictions: $f_2^{p-n}=-0.024 \pm 0.012$ 
\cite{MANK} and $f_2^{p-n}=-0.032 \pm 0.051$ 
\cite{Bal} (QCD sum rules), $f_2^{p-n}=0.028$ 
\cite{bag} (MIT bag model) and $f_2^{p-n}=-0.081$ 
\cite{inst} (instanton model).
The results can also be compared with values
obtained in analyses of the proton~\cite{HT_P} and neutron~\cite{HT_N}
data separately. Naively subtracting $f_2^n$ from $f_2^p$ gives 
$0.01 \pm 0.08$, which is consistent within
uncertainties with the above values for $f_2^{p-n}$.
However, different $Q^2$ ranges were used in the proton and neutron
fits, and different low-$x$ extrapolations implemented.

The larger uncertainty on $f_2^{p-n}$ from the $Q^2 > 0.8$~GeV$^2$ analysis
reflects the larger values of $Q_{\rm min}^2$ used here compared with
that used in the neutron analysis~\cite{HT_N}.
Fitting the neutron data from $Q^2 = 1$~GeV$^2$ rather from
$Q^2 = 0.5$~GeV$^2$ as in Ref.~\cite{HT_N} would increase the uncertainty
on $f_2^n$ appreciably, which, when combined with the proton data
fitted over the same range, would be more compatible with the
uncertainty from the present combined analysis.
This issue could be ameliorated with better quality data at higher
$Q^2$ ($Q^2 > 1$~GeV$^2$).
Data in this region on the proton and deuteron collected in Hall B 
at Jefferson Lab are presently being 
analysed.
Plans for high-precision measurements of the proton and neutron
structure functions at higher $Q^2$ are
also included in the 12 GeV energy upgrade of Jefferson Lab
\cite{12GeV}.

To summarize, we have presented an extraction of the Bjorken sum
in the $0.17 < Q^2 < 1.10$~GeV$^2$ range.
Being a nonsinglet quantity, the Bjorken sum simplifies the theoretical
analyses at both high $Q^2$ (using the OPE) and at low $Q^2$ (using $\chi$PT).
It thus provides us with a unique opportunity to understand better
the transition from perturbative QCD to the confinement region.
Combining with data at higher $Q^2$, we have extracted
the higher twist contributions to the sum.
We find $f_2^{p-n}$ small and the total higher twist contribution, 
for twists lower than eight, compatible with zero. 

This work was supported by the U.S. Department of Energy (DOE) and the U.S.
National Science Foundation. The Southeastern Universities Research 
Association operates the Thomas Jefferson National Accelerator 
Facility for the DOE under contract DE-AC05-84ER40150.

\vskip .1truein


\end{document}